# A perturbative quantized twist embedded in Minkowski spacetime

**J Strohaber**

Department of Physics, Florida A&M University, Tallahassee, Florida, 32307, USA

E-mail: jstroha1@gmail.com

**Abstract.** In this article, spatially-structured gravitational waves within the paraxial approximation are investigated. Drawing upon analogies between electrodynamics and general relativity, gauge invariant paraxial "electric" and "magnetic" parts of the Weyl conformal tensor are derived. In this approximation, a new gauge, which we call the *paraxial-traceless gauge*, is found. The polarization and degrees of freedom are investigated and compared with the Eardley-Newman-Penrose classification. Paraxial gravitational waves are found to share similarities with their electromagnetic counterparts in that they can possess a quantized amount of orbital angular momentum that can be transferred to matter.

## 1. Introduction

Shortly after Einstein discovered general relativity, he showed that linearization of the field equations predicted the existence of gravitational waves [1—4]. Gravitational waves are similar in many respects to electromagnetic waves in that they are wavelike disturbances carrying energy and momentum and traveling at the speed of light [3—6]. In 1993, the Nobel Prize in physics was awarded to Hulse and Taylor for measuring the orbital decay of PSR B1913+16 [7, 8]. The energy loss in the orbital decay of this binary system was found to be consistent with the energy loss predicted by general relativity. Until recently, this discovery has provided the strongest evidence for the existence of gravitational waves. In 2016, the LIGO and Virgo collaboration reported measuring a signal sweep of 35 to 250 Hz (up-chirp) [9] from what is thought to be gravitational radiation from two merging black holes [1]. Aside from a significance of greater than $5.1\sigma$ in their measurements, strong evidence of a genuine detection was found in the inversion of one signal due to the relative orientation of the Livingston and Hanford detectors [1]. The detection of gravitational waves is significant because it opens up a new field of spectroscopy allowing for the investigation of cosmological phenomena such as black-hole mergers [10], extra dimensions [11], and alternative metric theories [12].

The material presented in this work draws upon physics from two different fields of study: laser physics and gravitational physics in the weak-field limit. Since the advent of the laser, significant progress has been made by researchers in the field of optics and these advances have led to the unprecedented precision control of the properties of electromagnetic radiation, and its interaction with matter [13—15]. A subfield of optics that has drawn the attention of many scientists is singular optics with the well-known example being the so-called optical vortices [16—18]. Optical vortices belong to a family of transverse beam modes often characterized by an azimuthal phase-ramp resembling a spiral staircase. This azimuthal phase-ramp surrounds a phase singularity of zero amplitude along the beam axis. A line integral of the gradient of the phase enclosing the singularity gives rise to the Berry phase $\oint \vec{\nabla}\varphi \cdot dl = m2\pi$, where the integer value $m = 0, \pm 1, \pm 2,...$ is the so-called topological charge.

Interest in optical vortices has focused on the external orbital angular momentum (OAM) carried by these beams and how this physical quantity can be transferred to matter. Recent work in nonlinear singular optics includes the coherent transfer of OAM to Raman sidebands [19—21], and to high harmonics [22, 23]. Optical vortices have also interested the gravitational wave community, and various schemes have been proposed to use them in the reduction of detector noise [24—26]. It is the goal of this work to determine if gravitational waves can possess orbital angular momentum, and to investigate their influence on matter. While emphasis is placed on gravito-optical vortices, our derivation is kept general to include other paraxial modes.

This paper is organized as follows. Section 2 provides a covariant derivation of spatially-structured, free-space electromagnetic waves presented in a manner that facilitates derivation of paraxial gravitation waves (PGWs). Section 3 provides a brief review of the transverse-traceless gauge formalism for gravitational waves. In section 4, the metric perturbation of paraxial gravitational waves is derived, and in sections 5—7 the polarization, energy and flux densities, and the transfer of orbital angular momentum to a model detector are investigated.

## 2. Covariant derivation of paraxial electromagnetic fields

In electrodynamics, the propagation of electromagnetic waves is described by the wave equation. In the covariant formalism, the scalar and vector potentials are combined into a single entity known as the four-potential $A^a = (A^0, A^x, A^y, A^z)$. Here the speed of light has been set to unity $c = 1$ [27]. In the Lorentz gauge $\partial_\mu A^\mu = 0$, the four-potential can be shown to satisfy the wave equation

$$\left(\frac{\partial^2}{\partial x^2} + \frac{\partial^2}{\partial y^2} + \frac{\partial^2}{\partial z^2} - \frac{\partial^2}{\partial t^2}\right) A^\alpha = 0, \qquad (1)$$

which can also be written in the more compact form $\partial_\lambda \partial^\lambda A^\alpha = 0$. The four-potential can be put into its covariant form by lowering its index using the Minkowski metric $\eta_{\alpha\beta} = \eta^{\alpha\beta}$, which is taken here to have the diagonal components $\text{diag}(1, -1, -1, -1)$.

To derive the paraxial version of equation (1), a general solution of the form $A^\alpha = \mathcal{A}^\alpha(x, y, z)e^{-ik_\sigma x^\sigma}$ is assumed, where $k^\alpha = (\omega, k^x, k^y, k^z)$ is the propagation four-vector. Specializing to a monochromatic wave propagating along the $z$-direction $k_\alpha = (\omega, 0, 0, -k)$ and taking derivatives with respect to time yields the Helmholtz equation [28]

$$\left(\frac{\partial^2}{\partial x^2} + \frac{\partial^2}{\partial y^2} + \frac{\partial^2}{\partial z^2} + k^2\right) \mathcal{A}^\alpha e^{ikz} = 0. \qquad (2)$$

By taking partial derivatives with respect to the spatial coordinates in equation (2), and by neglecting the second derivative in the amplitude along the propagation direction $\partial^2 \mathcal{A}^\alpha / \partial z^2$ results in the paraxial wave equation (PWE) [29],

$$\left(\frac{\partial^2}{\partial x^2} + \frac{\partial^2}{\partial y^2} + 2ik\frac{\partial}{\partial z}\right) \mathcal{A}^\alpha = 0. \qquad (3)$$

Here $k = 2\pi/\lambda$ is the propagation constant with $\lambda$ being the wavelength of the radiation. The paraxial approximation reflects small changes in the beam amplitude along the propagation direction, and this approximation holds for radiation having a waist larger than a few wavelengths.

Under appropriate transformations, it has been shown that the PWE in equation (3) is mathematically identical to the 2D quantum harmonic oscillator [30], and Boyer *et al.* have shown by the method of separation of variables that solutions of equation (3) are possible in Cartesian, and cylindrical polar and elliptical coordinates [28]. In elliptical coordinates, the solutions have been recently found by M. Bandres *et al.* and are the so-called Ince-Gaussian modes [31]. The Hermite-Gaussian (Cartesian coordinates) and Laguerre-Gaussian (polar coordinates) modes have been presented by a number of authors [32, 33]. Although the derivation presented herein is valid for each of the three families of beam modes, special attention is given to the Laguerre-Gaussian solutions,

$$\psi = \frac{w_0}{w}\left(\frac{\sqrt{2}r}{w}\right)^{|\ell|} L_\rho^\ell\left(\frac{2r^2}{w^2}\right)\exp\left[-\left(\frac{1}{w^2}+i\frac{k}{2R}\right)r^2\right]e^{i\ell\theta}e^{-i(2\rho+|\ell|+1)\arctan\left(\frac{z}{z_0}\right)}. \tag{4}$$

The four-potential in equation (3) has 4 unknown components. One degree of freedom can be spent by invoking the Lorentz condition

$$\partial_z A^z = -\partial_0 A^0 - \partial_x A^x - \partial_y A^y, \tag{5}$$

which leaves three unknowns. A gauge freedom permits the following transformation to be performed

$$A_\alpha^{new} = A_\alpha^{old} - \partial_\alpha \Lambda. \tag{6}$$

Here $\Lambda = f(x,y,z)\exp(-ik_\beta x^\beta)$ is a scalar function that can be used to fix the gauge. This is accomplished by contracting the four-potential in equation (6) with the four-velocity of a stationary observer $u^\alpha = (1,0,0,0)$ co-moving with the source of radiation and then setting the result to zero i.e., $A_\alpha^{new} u^\alpha = 0$. The scalar function can then be found in terms of the *old* four-potential,

$$\begin{aligned} A_0^{new} &= A_0^{old} - \partial_0 \Lambda \\ 0 &= A_0^{old} + ik\Lambda \\ \Lambda &= iA_0^{old}/k \end{aligned}. \tag{7}$$

This procedure uses up all gauge freedoms and sets the temporal component of the four-potential to zero, and results in equation (5) be equal to $\partial_z A^z = -\partial_x A^x - \partial_y A^y$. To solve for $A^z$, the Ansatz $A^\alpha = \mathcal{A}^\alpha(x,y,z)e^{i(kz-\omega t)}$ is used to evaluated the z-derivative,

$$\partial_z A^\alpha = e^{-ik_\alpha x^\alpha}\left(\partial_z \mathcal{A}^\alpha + ik\mathcal{A}^\alpha\right). \tag{8}$$

Substituting the paraxial wave equation $\partial_z \mathcal{A}^\alpha = -i\left(\partial^x \partial_x \mathcal{A}^\alpha + \partial^y \partial_y \mathcal{A}^\alpha\right)/2k$ in place of the first term in equation (8) yields

$$\partial_z A^\alpha = ike^{-ik_\alpha x^\alpha}\left[\mathcal{A}^\alpha - \frac{1}{2k^2}\left(\partial^x \partial_x \mathcal{A}^\alpha + \partial^y \partial_y \mathcal{A}^\alpha\right)\right]. \tag{9}$$

A perturbative quantized twist embedded in Minkowski spacetime

When terms of order $\mathcal{O}(\lambda^2)$ [in the brackets] are neglected, equation (9) can be approximated as $\partial_z A^\alpha \approx ikA^\alpha$ or more specifically $\partial_z A^z \approx ikA^z$. By inserting this result into equation (5), the four-potential of paraxial electromagnetic radiation is found to be

$$A_\alpha^P = \left[ 0, A_x, A_y, -i\frac{1}{k}\left(\partial_x A^x + \partial_y A^y\right) \right]. \tag{10}$$

The superscript $P$ on the four-potential denotes that this is the four-potential within the paraxial approximation.

The paraxial four-potential in equation (10) is not unique because an additional gauge transformation $A_\alpha^{new} = A_\alpha^{old} - \partial_\alpha \Lambda'$ (where $\Lambda' = g(x,y,z)e^{j(kz-\omega t)}$) can be made such that $A_z$ is zero and $A_0$ is nonzero [34]. For example, by letting $A_z^{new} = 0$, the old $z$-component is $A_z^{old} = \partial_z \Lambda'$. Since the old time component is $A_0^{old} = 0$, the new time component is $A_0^{new} = -\partial_0 \Lambda'$. Using the paraxial approximation for the $z$-derivative of the scalar function $\partial_z \Lambda \approx ik\Lambda$ and $A_z^{old} = \partial_z \Lambda'$, the new time component is $A_0^{new} \approx A_z^{old}$. The $x$ and $y$ components transform as $A_\perp^{new} = A_\perp^{old} - \partial_\perp \Lambda'$ with $\Lambda' = -iA_0^{new}/k$. Because the new time component is $A_0^{new} \propto 1/k$, we get $A_x^{new} \approx A_x^{old}$. A similar calculation results in $A_y^{new} \approx A_y^{old}$. For this transformation, the new four-potential is $A_a^{new} = (A_z^{old}, A_x^{old}, A_y^{old}, A_0^{old})$. It is for this reason; we seek a quantity invariant under gauge transformations.

The electromagnetic field is a gauge invariant quantity given by a second rank, antisymmetric tensor known as the electromagnetic field strength tensor $F^{\alpha\beta} = \partial^\alpha A^\beta - \partial^\beta A^\alpha$. Using the four-potential in equation (10) immediately gives $F_{0b} = -ikA_b$. The $z$-components of $F^{\alpha\beta}$ are calculated from $F_{az} = \partial_a A_z - \partial_z A_a$. Using $A_{a,z} \approx ikA_a$ from equation (9), and the $z$-component of the four-potential $A_z = -i(\partial_x A^x + \partial_y A^y)/k$ we get

$$F_{az} = -ik\left[ A_a + \frac{1}{k^2}\partial_a\left(\partial^x A_x + \partial^y A_y\right) \right] \approx -ikA_a, \tag{11}$$

where terms of second order $\mathcal{O}(\lambda^2)$ in the bracketed expression have been neglected. The components $F_{xy} = -F_{yx} = \partial_x A_y - \partial_y A_x$ are as advertised. In the paraxial approximation, the electromagnetic field strength tensor takes the form,

$$F_{ab}^P \approx ik\begin{pmatrix} 0 & -A_x & -A_y & -A_z \\ A_x & 0 & -i(\partial_x A_y - \partial_y A_x)/k & -A_x \\ A_y & i(\partial_x A_y - \partial_y A_x)/k & 0 & -A_y \\ A_z & A_x & A_y & 0 \end{pmatrix}. \tag{12}$$

A perturbative quantized twist embedded in Minkowski spacetime

Choosing the transverse components of the four-potential as $A_x = i\alpha E / k$ and $A_y = i\beta E / k$, and using the polarization helicity $\sigma_z = i(\alpha\beta^* - \alpha^*\beta)$, (where $\sigma_z = 0$ for linear polarized light and $\sigma_z = \pm 1$ for left-handed and right-handed polarized light [34]), gives

$$F_{ab}^P \approx E_0 \begin{pmatrix} 0 & \alpha & \beta & \frac{i}{k}(\alpha\partial_x + \beta\partial_y) \\ -\alpha & 0 & \frac{i}{k}(\beta\partial_x - \alpha\partial_y) & \alpha \\ -\beta & -\frac{i}{k}(\beta\partial_x - \alpha\partial_y) & 0 & \beta \\ -\frac{i}{k}(\alpha\partial_x + \beta\partial_y) & -\alpha & -\beta & 0 \end{pmatrix} \psi. \quad (13)$$

From the field strength tensor of equation (13), the vector components of the electric and magnetic fields are retrieved in the following way $E^\alpha = F^\alpha{}_\beta u^\beta$ and $B^\alpha = {}^*F^\alpha{}_\beta u^\beta$, where $u^a$ is taken to be the four-velocity of a stationary observer $u^\alpha = (1,0,0,0)$ [35]. The gauge invariant electric and magnetic fields are found in this manner are, therefore,

$$\vec{E}^P = E_0 \left[ \alpha \hat{e}_x + \beta \hat{e}_y + \frac{i}{k}\left(\alpha \frac{\partial}{\partial x} + \beta \frac{\partial}{\partial y}\right) \hat{e}_z \right] \psi$$
$$\vec{B}^P = -E_0 \left[ \beta \hat{e}_x - \alpha \hat{e}_y + \frac{i}{k}\left(\beta \frac{\partial}{\partial x} - \alpha \frac{\partial}{\partial y}\right) \hat{e}_z \right] \psi \quad (14)$$

**3. Transverse-traceless gauge**
In General Relativity, the field equations describing the gravitational field is nonlinear and challenging to solve. However, much has been gained in understanding general relativistic effects from the linearized equations. Among these physical effects are *frame dragging* and the gravitational properties of light [34, 36, 37]. To linearize the field equations of general relativity, a small perturbation $h_{ab}$ is added to a flat background Minkowski spacetime [2,3],

$$g_{ab} = \eta_{ab} + h_{ab}. \quad (15)$$

Since both $g_{ab}$ and $\eta_{ab}$ are symmetric, it follows that $h_{ab}$ is symmetric. Using the metric of equation (15) and to first order in the perturbation, the vacuum Einstein tensor is [2],

$$G_{ab} = \bar{h}^c{}_{a,bc} + \bar{h}^c{}_{b,ac} - \eta_{ab}\bar{h}^{cd}{}_{,cd} - \partial_c\partial^c \bar{h}_{ab} = 0. \quad (16)$$

Here $\bar{h}_{ab} = h_{ab} - \eta_{ab}h/2$ is known as the trace-reverse and has the property $\bar{h} = -h$. For gravitational waves, the metric perturbation must satisfy the d'Alembertian wave operator, which appears as the fourth term in equation (16). By setting $\bar{h}^b{}_{a,b} = 0$, the Einstein tensor reduces to the vacuum wave equation

$$\left(\frac{\partial^2}{\partial x^2} + \frac{\partial^2}{\partial y^2} + \frac{\partial^2}{\partial z^2} - \frac{\partial^2}{\partial t^2}\right)h_{\alpha\beta} = 0. \tag{17}$$

Specializing to a gravitational wave traveling along the $z$-direction $h_{ab} = \mathfrak{h}_{ab}(x,y,z)e^{i(kz-\omega t)}$ and following the same procedure as was done in the electromagnetic case, the general relativistic wave equation of equation (17) takes on the familiar paraxial form,

$$\left(\frac{\partial^2}{\partial x^2} + \frac{\partial^2}{\partial y^2} + 2ik\frac{\partial}{\partial z}\right)\overline{\mathfrak{h}}_{\alpha\beta} = 0. \tag{18}$$

In the paraxial approximation, it comes as no surprise that the components of the metric perturbation $\overline{\mathfrak{h}}_{\alpha\beta}$ have the same eigensolutions as those found in the electromagnetic case with a difference being that $\overline{\mathfrak{h}}_{\alpha\beta}$ is a tensor quantity instead of a vector quantity $A^\alpha$. For this reason, it is not obvious at this point what polarization states *paraxial* gravitational waves will have.

There are sixteen components of the metric perturbation compared with the four components of the four-potential. Because the metric perturbation is symmetric, the number of unknown components is reduced to ten $n(n-1)/2 + n$. For non-paraxial gravitational waves [equation (17)], further restrictions, namely the four transverse conditions $k^b h_{ab} = 0$ and the four components of a vector $\xi^\mu$ used to fix the gauge, reduce the number of unknown components to that of 2. This procedure is shown in the Appendix for paraxial radiation and results in using up all gauge freedom. The remaining two components are said to be physically meaningful and are the two polarization states of the radiation. The constraint equations $h^a{}_a = 0$, $h_{ab}U^a = 0$ and $k^b h_{ab} = 0$ give rise to the well-known transverse-traceless gauge $h_{ab}^{\text{TT}}$. For a gravitational wave propagating along the $z$-direction, the metric perturbation is,

$$h_{ab}^{\text{TT}} = \begin{pmatrix} 0 & 0 & 0 & 0 \\ 0 & \alpha & \beta & 0 \\ 0 & \beta & -\alpha & 0 \\ 0 & 0 & 0 & 0 \end{pmatrix}\psi e^{i(kz-\omega t)}. \tag{19}$$

Here we have chosen the transverse components of the four-potential as $h_{yy} = -h_{xx} = -\alpha\psi$ and $h_{xy} = h_{yx} = \beta\psi$ [38]. In common notation, these two polarization states are called *plus* ($h_+$) and *cross* ($h_\times$) polarization states for the $h_{xx}$ and $h_{xy}$ components respectively. As with electromagnetic radiation, gravitation radiation also has circular polarization states $\mathfrak{h}_{\alpha\beta}^R = (\mathfrak{h}_{\alpha\beta}^+ + i\mathfrak{h}_{\alpha\beta}^\times)/\sqrt{2}$ and $\mathfrak{h}_{\alpha\beta}^L = (\mathfrak{h}_{\alpha\beta}^+ - i\mathfrak{h}_{\alpha\beta}^\times)/\sqrt{2}$.

To illustrate gravitation waves, an arrangement of test masses is situated in a region of space, and the proper displacement relative to a reference mass is calculated [2]. The proper displacement is found from the geodesic deviation,

$$\frac{d^2}{dt^2}S^a = R^a{}_{bcd}U^b U^c S^d. \tag{20}$$

A perturbative quantized twist embedded in Minkowski spacetime

The four-velocity is chosen to be those of an observer moving with the radiation source $U^\alpha = (1,0,0,0)$ and the proper time is $\tau \approx t$ [3]. The proper displacement along with $h_{ab} = \mathfrak{h}_{ab} e^{i(kz-\omega t)}$ can be approximated using a Dyson-type series,

$$S^\alpha \approx S^\alpha(0) + \frac{1}{2}\eta^{\alpha\beta}\mathfrak{h}_{\beta\gamma}S^\gamma e^{i(kz-\omega t)}. \tag{21}$$

The effects of the disturbances given by equation (21) on a spherical mass distribution are shown in figure 1. Additionally, our form of the metric perturbation $h_{ab}$ allows the polarization state to be conveniently defined by the helicity parameter $\sigma_z$.

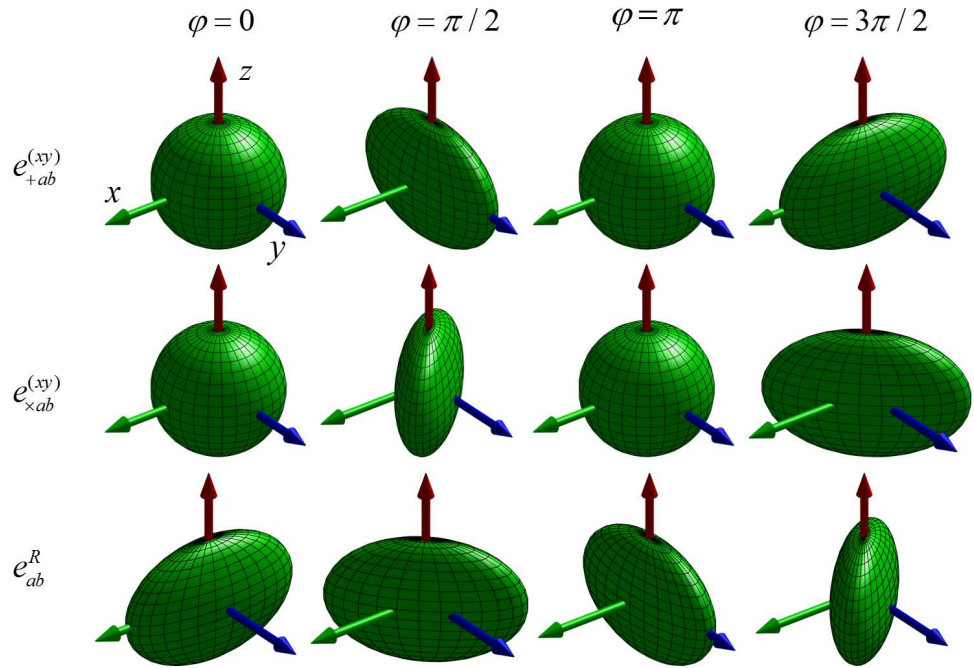

Figure 1. Spherical mass distributions illustrating polarization modes of gravitational waves in the TT-gauge. Rows (1) and (2) are for plus and cross polarizations, and row (3) is for right circularly polarized waves.

**4. Paraxial gravitational radiation**
In section 3, plane gravitational waves, formulated in terms of the metric perturbation, were briefly reviewed. The transverse-traceless condition required that the field polarization is transverse to the propagation vector $k_a \mathfrak{h}^{\alpha\beta} = 0$. For paraxial electromagnetic waves, we have shown that finite-sized beams possess longitudinal polarization [39], and for this reason, we expect a deviation from the TT-gauge. Three questions that naturally arise are (i) what type(s) of polarization states other than $h_+$ and $h_\times$ are found to contribute for paraxial gravitational radiation since there is more than one possible longitudinal polarization state as compared in the electromagnetic case, (ii) if other polarizations are found, how many degrees of freedom remain, and (iii) which metric theories support the various polarization states?

# A perturbative quantized twist embedded in Minkowski spacetime

To determine the metric perturbation associated with paraxial gravitational radiation, we following the program outlined for electromagnetic radiation in section 2. Beginning with a general metric perturbation

$$\bar{h}_{ab} = \begin{pmatrix} \bar{h}_{00} & \bar{h}_{01} & \bar{h}_{02} & \bar{h}_{03} \\ \bar{h}_{10} & \bar{h}_{11} & \bar{h}_{12} & \bar{h}_{13} \\ \bar{h}_{20} & \bar{h}_{21} & \bar{h}_{22} & \bar{h}_{23} \\ \bar{h}_{30} & \bar{h}_{31} & \bar{h}_{32} & \bar{h}_{33} \end{pmatrix} \qquad (22)$$

and using the Hilbert ("Lorentz") gauge condition $\partial_a \bar{h}^{ab} = 0$ along with $h^{ab} = \mathfrak{h}^{ab}(x,y,z) e^{-ik_\alpha x^\alpha}$ we find

$$\partial_a h^{ab} = e^{-ik_\alpha x^\alpha} \left( \partial_a \mathfrak{h}^{ab} - ik_a \mathfrak{h}^{ab} \right) = 0, \qquad (23)$$

which implies that $ik_a \mathfrak{h}^{\alpha\beta} = \partial_\alpha \mathfrak{h}^{\alpha\beta}$. This expression immediately demonstrates that the metric perturbation is not a pure transverse field since $\partial_\alpha \mathfrak{h}^{\alpha\beta}$ is not zero everywhere. Two constraint equations justified in the Appendix that will be used are $\mathfrak{h}^{ab} U_a = 0$, and the traceless condition $\mathfrak{h}^\alpha{}_\alpha = 0$. The first of these constraints can be used to set the time components to zero $\mathfrak{h}^{tb} = 0$. Next, we separate $ik_a \mathfrak{h}^{\alpha\beta} = \partial_\alpha \mathfrak{h}^{\alpha\beta}$ into time and space components

$$ik_t \mathfrak{h}^{tb} + ik_i \mathfrak{h}^{ib} = \partial_t \mathfrak{h}^{tb} + \partial_i \mathfrak{h}^{ib}. \qquad (24)$$

By making use of $\mathfrak{h}^{tb} = 0$, we find the relation $ik_i \mathfrak{h}^{ib} = \partial_i \mathfrak{h}^{ib}$. For a gravitational wave propagating along the $z$-direction, this last expression yields

$$ik\mathfrak{h}^{zb} = -\partial_x \mathfrak{h}^{xb} - \partial_y \mathfrak{h}^{yb} - \partial_z \mathfrak{h}^{zb}. \qquad (25)$$

The $\mathfrak{h}^{zb}$ components can be found by substituting the paraxial wave equation $\partial_z \mathfrak{h}^{ab} = -i \left( \partial^x \partial_x \mathfrak{h}^{ab} + \partial^y \partial_y \mathfrak{h}^{ab} \right) / 2k$ into equation (25)

$$\mathfrak{h}^{zb} = i \frac{1}{k} \left( \partial_x \mathfrak{h}^{xb} + \partial_y \mathfrak{h}^{yb} \right) + \frac{1}{2k^2} \left( \partial^x \partial_x \mathfrak{h}^{zb} + \partial^y \partial_y \mathfrak{h}^{zb} \right). \qquad (26)$$

By retaining terms of first order in the wavelength $\mathcal{O}(\lambda)$, the $z$-components of the metric perturbation are found to be

$$\begin{aligned} \mathfrak{h}^{zx} &= i \frac{1}{k} \left( \partial_x \mathfrak{h}^{xx} + \partial_y \mathfrak{h}^{yx} \right) \\ \mathfrak{h}^{zy} &= i \frac{1}{k} \left( \partial_x \mathfrak{h}^{xy} + \partial_y \mathfrak{h}^{yy} \right). \\ \mathfrak{h}^{zz} &= i \frac{1}{k} \left( \partial_x \mathfrak{h}^{xz} + \partial_y \mathfrak{h}^{yz} \right) \end{aligned} \qquad (27)$$

The first two equations for $\mathfrak{h}^{zx}$ and $\mathfrak{h}^{zy}$ can be substituted into the last equation for $\mathfrak{h}^{zz}$ to give

$$\mathfrak{h}^{zz} \approx -\frac{1}{k^2}\left[\partial_x\left(\partial_x\mathfrak{h}^{xx} + \partial_y\mathfrak{h}^{yx}\right) + \partial_y\left(\partial_x\mathfrak{h}^{xy} + \partial_y\mathfrak{h}^{yy}\right)\right], \quad (28)$$

which is of second order $\mathcal{O}(\lambda^2)$, and in our approximate is zero $h^{zz} \approx 0$. Finally, the traceless condition $h^{tt} - h^{xx} - h^{yy} - h^{zz} = 0$ can be used to further restrict the number of degrees of freedom. We have already set $h^{t\beta} = 0$, which implies that $h^{tt} = 0$, and equation (28) sets $h^{zz} \approx 0$; therefore, the traceless condition requires that $h_{yy} = -h_{xx}$. Symmetry further requires that $h_{xy} = h_{yx}$. The metric perturbation for paraxial gravitational radiation is therefore

$$h_{ab}^{PT} \approx \begin{pmatrix} 0 & 0 & 0 & 0 \\ 0 & h_{xx} & h_{xy} & \frac{i}{k}\left(\partial_x h_{xx} + \partial_y h_{xy}\right) \\ 0 & h_{xy} & -h_{xx} & \frac{i}{k}\left(\partial_x h_{xy} - \partial_y h_{xx}\right) \\ 0 & \frac{i}{k}\left(\partial_x h_{xx} + \partial_y h_{xy}\right) & \frac{i}{k}\left(\partial_x h_{xy} - \partial_y h_{xx}\right) & 0 \end{pmatrix}. \quad (29)$$

In equation (29), the superscript PT is used to indicate that the metric perturbation is for paraxial gravitational waves. Compared with the metric perturbation of equation (19), the metric perturbation of equation (29) appears to contain two additional polarization components.

We are not yet in a position to answer the questions posed earlier. Similar to the electromagnetic case, a gauge transformation of the form $\bar{h}_{\mu\nu}^{new} = \bar{h}_{\mu\nu}^{old} - \partial_\mu \xi_\nu - \partial_\nu \xi_\mu + \eta_{\mu\nu}\partial_\lambda \xi^\lambda$ can be made such that the $\bar{h}_{\alpha z}^{new}$ components are zero and for this reason, a gauge invariant quantity is sought. It is known that the Riemann tensor can be decomposed into the sum of three tensors [40],

$$R_{abcd} = C_{abcd} + E_{abcd} + S_{abcd}. \quad (30)$$

The first tensor on the right-hand side of equation (30) is the Weyl conformal tensor $C_{abcd}$, and the remaining two terms are the Einstein $E_{abcd}$ and the scalar $S_{abcd}$ tensors. The Weyl tensor is of interest here since it can be split into gauge invariant "electric" and "magnetic" parts by contracting $C_{abcd}$ and its dual ${}^*C_{abcd} = \varepsilon_{abef}C^{ef}{}_{cd}/2$ with time-like four-velocities similar to that done for the electromagnetic field strength tensor [5]. In vacuum, the Einstein and the scalar tensors vanish, leaving the Riemann tensor being equal to the Weyl tensor $R_{abcd} = C_{abcd}$. The electric and magnetic parts are symmetric, trace-free tensors that are invariant under gauge transformations. The electric and magnetic parts in tensor form are $E_{ab} = R_{acbd}U^cU^d$ and $B_{ab} = {}^*R_{acbd}U^cU^d$ respectively, and in matrix form they are written as,

$$E_{ab} = \begin{pmatrix} R_{xtxt} & R_{xtyt} & R_{xtzt} \\ R_{ytxt} & R_{ytyt} & R_{ytzt} \\ R_{ztxt} & R_{ztyt} & R_{ztzt} \end{pmatrix} \quad \text{and} \quad B_{ab} = \begin{pmatrix} R_{txyz} & R_{txzx} & R_{txxy} \\ R_{tyyz} & R_{tyzx} & R_{tyxy} \\ R_{tzyz} & R_{tzzx} & R_{tzxy} \end{pmatrix}. \quad (31)$$

A perturbative quantized twist embedded in Minkowski spacetime

Here $2R_{abcd} = h_{bc,ad} + h_{ad,bc} - h_{ac,bd} - h_{bd,ac}$ is the linearized Riemann tensor. When calculating the electric part, the Riemann tensor reduces to $2R_{a0c0} = k^2 h_{ac}$, and the electric part follows directly from the spatial part of the metric perturbation in equation (29),

$$E_{ab}^{PT} = \begin{pmatrix} h_{xx} & h_{xy} & \frac{i}{k}(\partial_x h_{xx} + \partial_y h_{xy}) \\ h_{xy} & -h_{xx} & \frac{i}{k}(\partial_x h_{xy} - \partial_y h_{xx}) \\ \frac{i}{k}(\partial_x h_{xx} + \partial_y h_{xy}) & \frac{i}{k}(\partial_x h_{xy} - \partial_y h_{xx}) & 0 \end{pmatrix}. \qquad (32)$$

The magnetic part is more easily determined using the general relativistic version of the Maxwell-Faraday equation [5]

$$-\frac{\partial B_{ij}}{\partial t} = \varepsilon_{jkm} \partial_k E_{im}. \qquad (33)$$

Keeping only terms of first order, the longitudinal components of the magnetic part $B_{iz}$ are

$$B_{iz} = -\frac{i}{k}(\partial^x E_{iy} - \partial^y E_{ix}). \qquad (34)$$

The remaining components are found in a similar fashion. For example, the $xx$-component is

$$B_{xx} = \frac{1}{k^2}(\partial_{xy} \mathfrak{h}^{xx} + \partial_{yy} \mathfrak{h}^{yx}) + i\frac{1}{k}\partial^z h_{xy}. \qquad (35)$$

Upon neglecting terms of order $\mathcal{O}(\lambda^2)$ and using the approximation $\partial_z \mathfrak{h}^{ab} = ik\mathfrak{h}^{ab}$, equation (35) reduces to $B_{xx} = -h_{xy}$. Continuing in this manner, the magnetic part of the Weyl (Riemann) tensor is found to be

$$B_{ab}^{PT} = \begin{pmatrix} h_{xy} & -h_{xx} & i\frac{1}{k}(\partial_x h_{xy} - \partial_y h_{xx}) \\ -h_{xx} & -h_{yx} & -i\frac{1}{k}(\partial_x h_{xx} + \partial_y h_{xy}) \\ i\frac{1}{k}(\partial_x h_{xy} - \partial_y h_{xx}) & -i\frac{1}{k}(\partial_x h_{xx} + \partial_y h_{xy}) & 0 \end{pmatrix}. \qquad (36)$$

Both electric and magnetic parts of the Riemann tensor are seen to contain 'longitudinal' components not found in the well-known TT-gauge. We have placed the superscript PT on these quantities to denote that they are field quantities derived in the paraxial approximation. The notation is also a reminder that the polarization of paraxial gravitational fields is not purely transverse $k_i \mathfrak{h}^{ib} \neq 0$ and that the traceless condition has been used.

**5. Polarization modes**
The electric and magnetic parts given by equations (32) and (36) are 3 by 3 symmetric, traceless tensors [5, 40, 42]. These properties restrict the number of possible basis tensors to a total of six. There exist more than one possible set of basis tensors; however, six basis tensors or polarization modes have been given by Eardley et al., and expressed in terms of the Newman-Penrose (NP) functions [12],

$$\Psi_2 = \frac{1}{6} R_{0z0z}$$
$$\Psi_3 = \frac{1}{2}\left(R_{0x0z} + iR_{0y0z}\right)$$
$$\Psi_4 = R_{0x0z} - R_{0y0y} + 2iR_{0x0y}$$
$$\Phi_{22} = R_{0x0x} + R_{0y0y}$$
(37)

Equations (37) are NP functions for GWs traveling along the $z$-direction. The real and imaginary parts of $\Psi_4$ give the plus $e^{(xy)}_{+\alpha\beta}$ and cross $e^{(xy)}_{\times\alpha\beta}$ polarizations found in the TT-gauge. The real and imaginary parts of $\Psi_3$ give $e^{(xz)}_{\times\alpha\beta}$ and $e^{(yz)}_{\times\alpha\beta}$ respectively, and the functions $\Psi_2$ and $\Phi_{22}$ give the polarization modes $e^{(zz)}_{\|\alpha\beta}$ and $e^{(xy)}_{\odot\alpha\beta}$. In matrix form, these polarization modes are

$$e^{(xy)}_{+\alpha\beta} = \begin{pmatrix} 1 & 0 & 0 \\ 0 & -1 & 0 \\ 0 & 0 & 0 \end{pmatrix} \quad e^{(xy)}_{\times\alpha\beta} = \begin{pmatrix} 0 & 1 & 0 \\ 1 & 0 & 0 \\ 0 & 0 & 0 \end{pmatrix} \quad e^{(xz)}_{\times\alpha\beta} = \begin{pmatrix} 0 & 0 & 1 \\ 0 & 0 & 0 \\ 1 & 0 & 0 \end{pmatrix}$$
$$e^{(yz)}_{\times\alpha\beta} = \begin{pmatrix} 0 & 0 & 0 \\ 0 & 0 & 1 \\ 0 & 1 & 0 \end{pmatrix} \quad e^{(z)}_{\|\alpha\beta} = \begin{pmatrix} 0 & 0 & 0 \\ 0 & 0 & 0 \\ 0 & 0 & 1 \end{pmatrix} \quad e^{(xy)}_{\odot\alpha\beta} = \begin{pmatrix} 1 & 0 & 0 \\ 0 & 1 & 0 \\ 0 & 0 & 0 \end{pmatrix}$$
(38)

In our notation, $+(\times)$ denotes plus (cross) polarization in a plane indicated by the coordinate(s) in parenthesis ( ). The $e^{(z)}_{\|\alpha\beta}$ polarization tensor corresponds to a 'stretching' mode along the $z$-direction, and that of $e^{(xy)}_{\odot\alpha\beta}$ corresponds to a symmetric 'breathing' mode in the $xy$-plane. The modes $e^{(xy)}_{+\alpha\beta}$ and $e^{(xy)}_{\times\alpha\beta}$ are called tensor modes since a 180° rotation around the $z$-axis results in a symmetry operation. The modes $e^{(xz)}_{\times\alpha\beta}$ and $e^{(yz)}_{\times\alpha\beta}$ are called vector modes since a symmetry occurs for a rotation of 360°, and the modes $e^{(zz)}_{\|\alpha\beta}$, and $e^{(xy)}_{\odot\alpha\beta}$ are scalar mode since they are rotationally symmetric about the $z$-axis. These mode helicities are denoted by $s = \pm2, \pm1, 0$, and are linked to the spin of massless particles associated with the gravitational radiation. For paraxial gravitation waves, it is found that the waves consist of 2 tensor modes of spin 2, and 2 vector modes of spin 1. The evolution of these polarization states is shown in figure 2.

We are now in a position to address the questions posed earlier. From equations (32), (36) and (38), it is found that PGWs, to first order in $\lambda$, have two additional polarization modes, which appear in the NP-Eardley classification. By inspecting the components of the electric and magnetic parts, the two additional polarization modes $e^{(xz)}_{\times\alpha\beta}$ and $e^{(yz)}_{\times\alpha\beta}$ are seen to depend completely on the two degrees of freedom $h_{xx}$ and $h_{xy}$. For this reason, paraxial GW do not result in additional degrees of freedom.

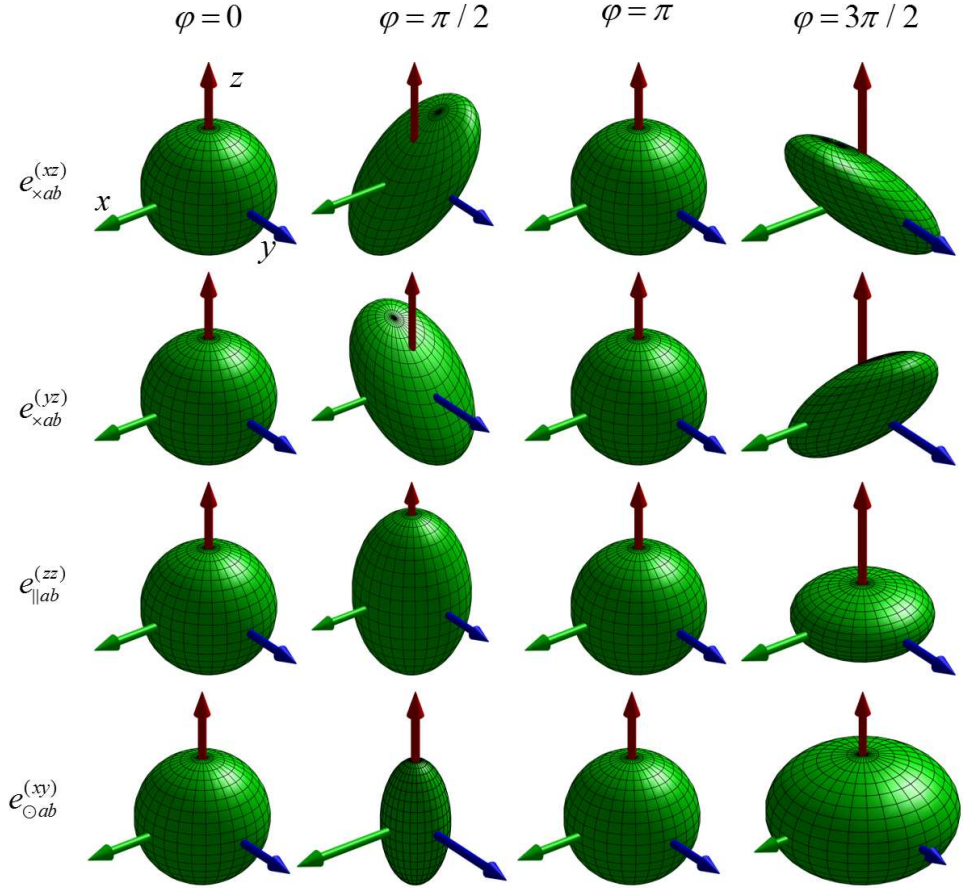

**Figure 2.** Evolution of polarization modes beyond the TT-gauge. Rows (1) and (2) are "longitudinal" vector modes in the $xz$ and $yz$ planes respectively. Rows (3) and (4) are scalar polarization modes with $e^{(zz)}_{\|ab}$ being a longitudinal "stretching" mode and $e^{(xy)}_{\odot ab}$ a "breathing" mode in the $xy$ plane. For paraxial gravitational wave to first order in $\lambda$, the vector modes in rows 1 and 2 are found.

In testing for permitted GWs from various viable metric theories, Eardley *et al.*, based their classification scheme on the polarization modes. According to their scheme, our paraxial GWs appear to fit into Class III$_5$. In this class, all standard observers will measure the absence of $e^{(z)}_{\|ab}$ and the presence of $e^{(xz)}_{\times ab}$ and $e^{(yz)}_{\times ab}$ with the presence of modes $e^{(xy)}_{+ab}$, $e^{(xy)}_{\times ab}$ and $e^{(xy)}_{\odot ab}$ being observer dependent. The classification III$_5$ belongs to a more general class than that permitted by general relativity, but it is allowed by other viable metric theories [12]. The reason for this is that general relativity restricts extra degrees of freedom [2], which is not the case for paraxial gravitational waves. As a note, when higher order terms $\mathcal{O}(\lambda^2)$ are kept in our derivations, we find evidence of $e^{(z)}_{\|\alpha\beta}$ and $e^{(xy)}_{\odot\alpha\beta}$ polarization modes in addition to an increase in the degrees of freedom to 3.

The additional polarization modes can be shown to be locally observer dependent. For paraxial electromagnetic waves, the longitudinal component of the polarization $A_z$ is known to be due to wavefront curvature [39]. To investigate the $E^{PT}_{az}$ components of PT-waves, let the normal to the plane of a detector measuring a TT-wave $E_{ab}^{TT}$ propagating along the z-axis vary by the angle $\theta$ in the xz-plane and by the angle $\varphi$ in the yz-plane. Using the notation $C_x = \cos(x)$ and $S_x = \sin(x)$, the resulting transformation is

$$R^\dagger E_{ab}^{TT} R = \begin{pmatrix} \alpha C_\theta^2 & \beta C_\theta C_\varphi & -\alpha C_\theta S_\theta - \beta C_\theta S_\varphi \\ \beta C_\varphi C_\theta & -\alpha C_\varphi^2 & -\beta C_\varphi S_\theta + \alpha C_\varphi S_\varphi \\ -\alpha C_\theta S_\theta - \beta S_\varphi C_\theta & -\beta S_\theta C_\varphi + \alpha C_\varphi S_\varphi & \alpha S_\theta^2 + 2\beta S_\varphi S_\theta - \alpha S_\varphi^2 \end{pmatrix}. \quad (39)$$

For small angles, the approximations $\cos(x) \approx 1$, $\sin(x) \approx x$ and $\sin^2(x) \approx 0$ simplifies equation 39 for $R^\dagger E_{ab}^{TT} R$ to

$$R^\dagger E_{ab}^{TT} R \approx \begin{pmatrix} \alpha & \beta & -\alpha\theta - \beta\varphi \\ \beta & -\alpha & -\beta\theta + \alpha\varphi \\ -\alpha\theta - \beta\varphi & -\beta\theta + \alpha\varphi & 0 \end{pmatrix} \quad (40)$$

For the lowest order Gaussian mode, the derivatives in equation (32) are proportional to the spatial coordinates i.e., $\partial_x \mathfrak{h}_{ab} \propto x \mathfrak{h}_{ab} / w_0^2$ and $\partial_x \mathfrak{h}_{ab} \propto y \mathfrak{h}_{ab} / w_0^2$. The wavefront curvature is small and requires an incremental observation angle. It can be shown that the angle, relative to the z-axis, of a vector normal to the wavefront can be approximated as $\theta \approx r / k w_0^2$. Comparing equation (40) with equation (32), and the $E_{zz}$ component of equation (39) with equation (28) nicely demonstrates that the origin of the observer dependent polarization modes are due to wavefront curvature.

Lastly, the paraxial 'electromagnetic' gravitational fields are expressed using the polarizations tensors $e_{+ab}^{(xy)}$, $e_{\times ab}^{(xy)}$, $e_{\times ab}^{(xz)}$, $e_{\times ab}^{(yz)}$, $e_{\|ab}^{(zz)}$, and $e_{\odot ab}^{(xy)}$ of equation (38) for comparison with their electromagnetic counterparts given in equation (14),

$$\begin{aligned} E_{\alpha\beta}^{PT} &= \left[ \alpha e_{+\alpha\beta}^{(xy)} + \beta e_{\times\alpha\beta}^{(xy)} + i\frac{1}{k}\left(\alpha\frac{\partial}{\partial x} + \beta\frac{\partial}{\partial y}\right)e_{\times\alpha\beta}^{(xz)} + i\frac{1}{k}\left(\beta\frac{\partial}{\partial x} - \alpha\frac{\partial}{\partial y}\right)e_{\times\alpha\beta}^{(yz)} \right]\psi \\ B_{\alpha\beta}^{PT} &= \left[ \beta e_{+\alpha\beta}^{(xy)} - \alpha e_{\times\alpha\beta}^{(xy)} + i\frac{1}{k}\left(\beta\frac{\partial}{\partial x} - \alpha\frac{\partial}{\partial y}\right)e_{\times\alpha\beta}^{(xz)} - i\frac{1}{k}\left(\alpha\frac{\partial}{\partial x} + \beta\frac{\partial}{\partial y}\right)e_{\times\alpha\beta}^{(yz)} \right]\psi \end{aligned}. \quad (41)$$

### 6. Energy density and Poynting vector

Two important physical quantities associated with radiation are the energy and flux densities (Poynting vector). In electrodynamics, these quantities are given by the energy-momentum tensor $T^{\mu\nu} = \left(F^{\mu\alpha}F^\nu{}_\beta - \eta^{\mu\nu}F_{\alpha\beta}F^{\alpha\beta}/4\right)/\mu_0$ [27]. The energy density is given by $T^{00}$, and the flux density is found from $P^a = c(T^{0x}, T^{0y}, T^{0z})$. The gravitational analog of the energy-momentum tensor is the Bell-Robertson tensor or the so-called super-energy-momentum tensor [5],

$$T_{ab}{}^{cd} = \frac{1}{4}\left(R_{a\mu b\nu}R^{c\mu d\nu} + {}^*R_{a\mu b\nu}{}^*R^{c\mu d\nu}\right). \quad (42)$$

From equation (42), the super-energy density and super-Poynting vector are found by contraction with time-like velocity vectors

A perturbative quantized twist embedded in Minkowski spacetime

$$T_{tttt} = \frac{1}{4}\left(E^*_{ab}E^{ab} + B^*_{ab}B^{ab}\right), \qquad (43.1)$$

$$T_{attt} = \frac{1}{4}\varepsilon_{abc}\left(B^*_{xb}E^{xc} + B^*_{yb}E^{yc} + B^*_{zb}E^{zc}\right). \qquad (43.2)$$

As a specific example, these quantities are calculated for a gravitational Laguerre-Gaussian $LG_0^\ell$ wave

$$T_{tttt} \approx |\psi|^2, \qquad (44.1)$$

$$T_{attt} \approx \left[\frac{r}{R(z)}\hat{e}_r + \frac{1}{k}\left(\frac{\ell}{r} - \sigma_z\frac{1}{2}\frac{\partial}{\partial r}\right)\hat{e}_\theta + \hat{e}_z\right]|\psi|^2. \qquad (44.2)$$

These energy and flux densities are similar in form to the energy density and Poynting vector found for paraxial electromagnetic beams [34]. For *gravito-optical vortices*, the energy flux follows integral curves that 'spiral' around the gravito-optical axis which is again similar to that found for electromagnetic vortices.

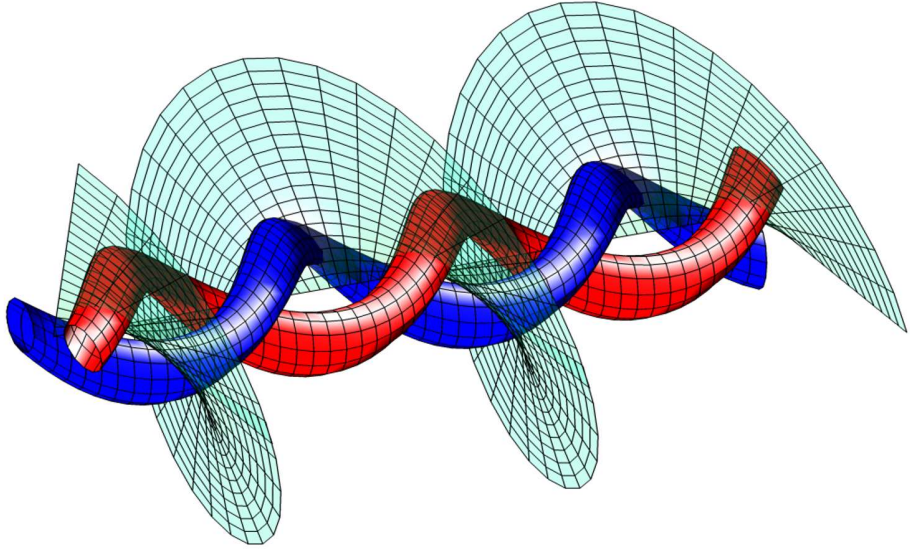

Figure 3. Illustration of the wavefront and Poynting vector of a gravito-optical Laguerre-Gaussian mode $LG_0^1$. For illustration purposes, the amplitude, wavefront curvature and Gouy phase have been neglected. The phase is shown by the spiral plane, and the flux density, Poynting vector, is represented by rings of particles forming the red and blue tubes.

The spiraling of the Poynting vector is shown in figure 3. The red and blue tubes are constructed of rings of particles in a plane in which the propagation vector is normal, and the centers of the rings follow integral curves. Each tube locally distorts according to the proper displacement given in equation (21) and depends on the local phase of the radiation. Longitudinal distortions have been neglected, and the amplitude has been taken to be uniform around the vicinity of the test particles. The red and blue colors of the tubes denote that they are out of phase with each other so as one contracts along a particular direction the other expands in that same direction. These tubes reside on opposites sides of the gravito-optical vortex.

**7. Force exerted on a spring-mass system**

Electromagnetic fields are known to impart linear momentum to and induce torques on material bodies [34, 43, 44]. To investigate the transfer of momentum from gravito-optical vortices to matter, a model spring-mass detector is constructed [4, 44]. The spring-mass system consists of two identical, slowly moving massive particles connected by a spring and separated by an incremental displacement $2\delta r^\alpha$. The spring force on each mass is given by $f$. The geodesic deviation of the two neighboring massive particles relative to the center of mass of the system is

$$\frac{d^2}{dt^2}\left(r^\alpha + \delta r^\alpha\right) = (r^\delta + \delta r^\delta)R^{\alpha+}{}_{00\delta} + f^\alpha/m \qquad (45)$$
$$\frac{d^2}{dt^2}\left(r^\alpha - \delta r^\alpha\right) = (r^\delta - \delta r^\delta)R^{\alpha-}{}_{00\delta} - f^\alpha/m .$$

Since the field $R_{abcd}$ spatially varies over the displacement of the masses, the field strength tensors in equation (45) are functions of position, and the following shorthand notation has been used $R^{\alpha+}{}_{00\delta} = R^\alpha{}_{00\delta}(r^\delta + \delta r^\delta)$ and $R^{\alpha-}{}_{00\delta} = R^\alpha{}_{00\delta}(r^\delta - \delta r^\delta)$. Adding equations (45), the acceleration of the center of mass is found to be

$$2\frac{d^2}{dt^2}r^\alpha = \left(R^{\alpha+}{}_{00\delta} + R^{\alpha-}{}_{00\delta}\right)r^\delta + \left(R^{\alpha+}{}_{00\delta} - R^{\alpha-}{}_{00\delta}\right)\delta x^\delta . \qquad (46)$$

For small displacements of the masses near the center of mass, the Riemann tensors can be expanded in a Taylor series around that point,

$$R^{\alpha+}{}_{\beta\gamma\delta}(r^\alpha + \delta x^\alpha) = R^\alpha{}_{\beta\gamma\delta}(r^\alpha) + R^\alpha{}_{\beta\gamma\delta,\sigma}(r^\alpha)\delta x^\sigma \qquad (47)$$
$$R^{\alpha-}{}_{\beta\gamma\delta}(r^\alpha - \delta x^\alpha) = R^\alpha{}_{\beta\gamma\delta}(r^\alpha) - R^\alpha{}_{\beta\gamma\delta,\sigma}(r^\alpha)\delta x^\sigma .$$

Adding and subtracting these field quantities results in $R^{\alpha+}{}_{\beta\gamma\delta} - R^{\alpha-}{}_{\beta\gamma\delta} = 2R^\alpha{}_{\beta\gamma\delta,\sigma}\delta x^\sigma$ and $R^{\alpha+}{}_{\beta\gamma\delta} + R^{\alpha-}{}_{\beta\gamma\delta} = 2R^\alpha{}_{\beta\gamma\delta}$. Substituting these two results into equation (46) yields

$$\frac{d^2}{dt^2}r^\alpha = R^\alpha{}_{00\delta}r^\delta + R^\alpha{}_{00\delta,\sigma}\delta x^\sigma \delta x^\delta . \qquad (48)$$

Using interchange symmetry $R^\alpha{}_{\beta\gamma\delta,\sigma} = R^\gamma{}_{\delta\alpha\beta,\sigma}$, and the Bianchi Identity $R^0{}_{\delta\alpha 0,\sigma} + R^0{}_{\delta\sigma\alpha,0} + R^0{}_{\delta 0\sigma,\alpha} = 0$ [2—4], the acceleration is found to be

$$\frac{d^2}{dt^2}r^\alpha = R^\alpha{}_{00\delta}r^\delta - R^0{}_{\delta 0\sigma,\alpha}\delta x^\sigma \delta x^\delta - R^0{}_{\delta\sigma\alpha,0}\delta x^\sigma \delta x^\delta . \qquad (49)$$

Because $R^0{}_{\delta\sigma\alpha,0}$ is periodic in time, it can be dropped since we are interested in the drift of the detector's center of mass over an extended period of time. Substituting the electric part $E_{ab}$ for the

A perturbative quantized twist embedded in Minkowski spacetime

Riemann tensor and using the notation $Q^{\sigma\delta} = \delta x^\sigma \delta x^\delta$, the acceleration of the center of mass of the detector is found to be

$$\frac{d^2}{dt^2}r^\alpha = E_{\alpha\delta}r^\delta - Q^{\sigma\delta}\frac{\partial}{\partial x^\alpha}E_{\sigma\delta}. \tag{50}$$

The first term in equation (50) is the geodesic deviation given in equation (20), and the second term gives rise to momentum transfer to the spring-mass system from the gravitational wave. In the case of a uniform *non-paraxial* plane gravitational wave, the second term is $Q^{\sigma\delta}\partial E_{\sigma\delta}/\partial z = ikQ^{\sigma\delta}E_{\sigma\delta}$, which yields a simple graviton momentum transfer picture along the propagation direction. For a *paraxial* gravito-optical vortex of the form $\mathrm{LG}_0^\ell$, the second term produces a torque about the gravitational wave axis equal to

$$\left(yQ^{\sigma\delta}\frac{\partial}{\partial x}E_{\sigma\delta} - xQ^{\sigma\delta}\frac{\partial}{\partial y}E_{\sigma\delta}\right)\hat{k} = \ell Q^{\sigma\delta}E_{\sigma\delta}\hat{k}. \tag{51}$$

Equation (51) shows that the torque is quantized by the azimuthal mode number of the $\mathrm{LG}_0^\ell$ mode, and the direction of the torque is found to depend on the sign of $\ell$. When the azimuthal mode number is zero, there is no resulting torque on the spring-mass system about the axis of the gravitational wave vortex.

**8. Conclusions**
Gravitational waves have been calculated within the paraxial approximation. The resulting waves are found to contain two additional polarization states allowed by the Newman-Penrose-Eardley classification of viable metric theories. The components of the additional polarization modes depend entirely on the 2 degrees of freedom found in the TT-gauge. As with electromagnetic vortices, gravito-optical vortices are found to possess a nonzero amount of orbital angular momentum which we have shown produces a quantized torque on a model detector.

**Appendix: gauge fixing**
A gauge transformation $\bar{h}_{\mu\nu}^{new} = \bar{h}_{\mu\nu}^{old} - \partial_\mu\xi_\nu - \partial_\nu\xi_\mu + \eta_{\mu\nu}\partial_\lambda\xi^\lambda$ can be used to set the time component $\bar{h}_{00} = 0$, and the trace $\bar{h}_\mu^\mu = 0$ to zero [4]. Using $h_{ab} = \mathfrak{h}_{ab}(x,y,z)e^{ik_\alpha x^\alpha}$ and the vector $\xi_\mu = B_\mu(x,y,z)e^{ik_\sigma x^\sigma}$, the gauge transformation yields

$$\mathfrak{h}_{\mu\nu}^{new}e^{ik_\alpha x^\alpha} = \mathfrak{h}_{\mu\nu}^{old}e^{ik_\alpha x^\alpha} - \partial_\mu\left(B_\nu e^{ik_\sigma x^\sigma}\right) - \partial_\nu\left(B_\mu e^{ik_\sigma x^\sigma}\right) + \eta_{\mu\nu}\partial_\lambda\left(B^\lambda e^{ik_\sigma x^\sigma}\right). \tag{A.1}$$

Taking the trace and setting it to zero $\bar{h}_\mu^\mu = 0$ and using $\eta^{\mu\nu}\eta_{\mu\nu} = 4$ yields after some algebra,

$$k_\lambda B^\lambda = i\frac{1}{2}\eta^{\mu\nu}\mathfrak{h}_{\mu\nu}^{old} + i\partial_\lambda B^\lambda. \tag{A.2}$$

Now setting $\mathfrak{h}_{0\nu}^{new} = 0$, we find the time component of the vector $B_\alpha$ that sets $\mathfrak{h}_{00}^{new} = 0$. This component is

A perturbative quantized twist embedded in Minkowski spacetime

$$B_0 = \frac{1}{2k}\left(i\mathfrak{h}_{00}^{old} - k_\lambda B^\lambda + i\partial_\lambda B^\lambda\right). \tag{A.3}$$

Combining equation (A.2) with equation (A.3), the time component of $B_a$ can be given in terms of the old components of the metric perturbation

$$B_0 = \frac{i}{2k}\left(\mathfrak{h}_{00}^{old} - \frac{1}{2}\eta^{\mu\nu}\mathfrak{h}_{\mu\nu}^{old}\right). \tag{A.4}$$

This is the same time component found in the TT-gauge. Spatial components of $B_a$ can be found by setting $\mathfrak{h}_{0i}^{new} = 0$ in equation (A.1) to give,

$$B_j = -\frac{i}{k}\left(\mathfrak{h}_{0j}^{old} - ikB_0 - \partial_j B_0\right). \tag{A.5}$$

Substituting equation (A.5) into equation (A.6) yields

$$B_j = -\frac{i}{2k}\left[2\mathfrak{h}_{0j}^{old} + \left(\mathfrak{h}_{00}^{old} - \frac{1}{2}\eta^{\mu\nu}\mathfrak{h}_{\mu\nu}^{old}\right) - i\partial_j\left(\mathfrak{h}_{00}^{old} - \frac{1}{2}\eta^{\mu\nu}\mathfrak{h}_{\mu\nu}^{old}\right)\right]. \tag{A.6}$$

The last term in the brackets is an extra term that is not found in the TT-gauge, and it is due to the fact that, in general, the field depends on the spatial coordinates.

A perturbative quantized twist embedded in Minkowski spacetime